\documentclass[journal]{IEEEtran}
\usepackage{makecell}
\usepackage{graphicx,amssymb,amsmath}
\usepackage{multicol}
\usepackage[noadjust]{cite}
\usepackage{setspace}
\usepackage{float}
\usepackage{amsthm,pifont}
\usepackage{cases,subeqnarray}
\usepackage{bm,multirow,bigstrut}
\usepackage{amsmath, amsthm, amssymb}
\usepackage{textcomp}
\usepackage{latexsym,bm}
\usepackage{booktabs}
\usepackage{mathtools}
\usepackage{dsfont}
\usepackage{extarrows}
\usepackage{epsfig}
\usepackage{epsfig}
\usepackage{epstopdf}
\usepackage[noend]{algpseudocode}
\usepackage{colortbl}
\usepackage{booktabs}
\usepackage{graphicx} 
\usepackage{array}    
\usepackage{booktabs}
\usepackage{enumitem}
\usepackage{diagbox}
\usepackage{tikz}
\usetikzlibrary{trees,shadows.blur}
\usepackage{forest}

\usepackage{lineno} 

\usepackage{amsmath,amsfonts}
\usepackage{array}
\usepackage{subcaption}
\usepackage{textcomp}
\usepackage{stfloats}
\usepackage{url}
\usepackage{verbatim}
\usepackage{graphicx}
\usepackage{multirow}

\usepackage{cite}

\usepackage[bookmarksnumbered=true]{hyperref}
\hypersetup{
hidelinks,
colorlinks=true,
linkcolor=blue,
citecolor=blue,
urlcolor=black
}
\usepackage[linesnumbered,ruled,vlined]{algorithm2e}
\usepackage{bbm} 

\usepackage{amsthm} 

\usepackage{hyperref}
\usepackage{colortbl}
\usepackage{xcolor}

\usepackage{enumerate} 

\begin{document}

\definecolor{headergray}{RGB}{220,220,220}
\definecolor{lightgray}{RGB}{245,245,245}
\definecolor{blue-green}{RGB}{252,41,30}

\newcounter{probcounter}
\newcommand{\refporbcounter}[1]{\refstepcounter{probcounter}\theprobcounter \label{#1}} 

\title{Chain-of-Thought for Large Language Model-empowered Wireless Communications}

\author{
Xudong Wang, 
Jian Zhu,
Ruichen Zhang,
Lei Feng,~\IEEEmembership{Member,~IEEE,}
Dusit Niyato,~\IEEEmembership{Fellow,~IEEE,} \\
Jiacheng Wang,
Hongyang Du,
Shiwen Mao,~\IEEEmembership{Fellow,~IEEE,}
and Zhu Han,~\IEEEmembership{Fellow,~IEEE,}

\thanks{\textit{Corresponding author: Lei Feng.}}
\thanks{
X. Wang, J. Zhu, and L. Feng are with the State Key Laboratory of Networking and Switching Technology, Beijing University of Posts and Telecommunications, Beijing, China, 100876 
(e-mail: xdwang@bupt.edu.cn, zhujian@bupt.edu.cn, fenglei@bupt.edu.cn).}
\thanks{R. Zhang, D. Niyato and J. Wang are with the College of Computing and Data Science,
Nanyang Technological University, Singapore (e-mail: ruichen.zhang@ntu.edu.sg, dniyato@ntu.edu.sg, jiacheng.wang@ntu.edu.sg).}
\thanks{H. Du is with the Department of Electrical and Electronic Engineering, University of Hong Kong, Pok Fu Lam, Hong Kong SAR, China (e-mail: duhy@eee.hku.hk).}
\thanks{S. Mao is with the Department of Electrical and Computer Engineering,
Auburn University, Auburn, AL 36849, USA (e-mail: smao@ieee.org).}
\thanks{Z. Han is with the University of Houston, Houston TX 77004, USA, and
also with the Department of Computer Science and Engineering, Kyung Hee
University, Seoul 446701, South Korea (e-mail: hanzhu22@gmail.com).}
}



\maketitle

\begin{abstract}
Recent advances in large language models (LLMs) have opened new possibilities for automated reasoning and decision-making in wireless networks. However, applying LLMs to wireless communications presents challenges such as limited capability in handling complex logic, generalization, and reasoning. Chain-of-Thought (CoT) prompting, which guides LLMs to generate explicit intermediate reasoning steps, has been shown to significantly improve LLM performance on complex tasks. Inspired by this, this paper explores the application potential of CoT-enhanced LLMs in wireless communications. Specifically, we first review the fundamental theory of CoT and summarize various types of CoT. We then survey key CoT and LLM techniques relevant to wireless communication and networking. Moreover, we introduce a multi-layer intent-driven CoT framework that bridges high-level user intent expressed in natural language with concrete wireless control actions. Our proposed framework sequentially parses and clusters intent, selects appropriate CoT reasoning modules via reinforcement learning, then generates interpretable control policies for system configuration. Using the unmanned aerial vehicle (UAV) network as a case study, we demonstrate that the proposed framework significantly outperforms a non-CoT baseline in both communication performance and quality of generated reasoning.
\end{abstract}


\section{Introduction}


Wireless communication systems have witnessed remarkable advancements over the past decade, driven by the increasing demand for higher data rates, lower latency, and improved energy efficiency. However, the growing complexity of wireless networks necessitates innovative approaches to network design, optimization, and decision-making. In this context, large language models (LLMs) have emerged as a powerful tool to address these challenges. LLMs, such as GPT and other transformer-based architectures, have demonstrated remarkable capabilities in various tasks beyond natural language processing (NLP), including network optimization, resource management, and adaptive decision-making~\cite{10681550}. By leveraging LLMs in wireless communications, the networks can achieve improved performance in dynamic environments, thereby enhancing spectrum efficiency, mitigating interference, and improving energy utilization.

Despite the promising potential of LLMs in wireless communications, several limitations hinder their widespread adoption and performance optimization. One significant challenge is the lack of interpretability and transparency in LLM-driven decisions. As black-box models, LLMs make it difficult for network operators to understand the rationale behind certain decisions, particularly in dynamic wireless environments with stringent performance requirements~\cite{10685369}. This opacity reduces their reliability in mission-critical scenarios. Moreover, conventional LLMs face difficulties with complex, multi-step reasoning tasks crucial for wireless network optimization, including interference management, beamforming, and spectrum allocation. Additionally, the computational overhead associated with LLM inference poses a challenge for latency-sensitive applications. Addressing these challenges is key to realizing the full promise of LLMs in wireless systems.

To overcome these challenges, Chain-of-Thought (CoT) has emerged as a powerful technique to enhance the reasoning capabilities of LLMs~\cite{wei2022chain}. CoT is a strategy that encourages the model to explicitly break down complex problems into sequential reasoning steps, thus improving the multi-step decision-making ability. Incorporating CoT technology enhances LLM interpretability by making reasoning transparent and helping network operators trace decisions, which is especially crucial for dynamic, context-aware optimization in wireless systems. Moreover, CoT also improves model accuracy and reliability in tasks requiring step-by-step deduction. By leveraging CoT, wireless networks can achieve more robust and efficient decision-making, ultimately enhancing network performance and resilience in complex and dynamic environments.

Considering the critical role of CoT in LLM-empowered reasoning, this paper explores its potential in wireless communications. First, we analyze the principles and benefits of CoT, while categorizing CoT techniques. Then we provide a comprehensive review of the practical applications of CoT-enhanced LLMs in wireless communications, such as path planning and semantic communication. Finally, we design a CoT-enhanced LLM framework for intent-driven wireless networks, and illustrate its effectiveness through a case study involving unmanned aerial vehicle (UAV) deployment and resource allocation in intent-aware low-altitude networks. The main contributions of this paper are as follows.

\begin{itemize}

    \item We provide a comprehensive overview of CoT principles and advantages, and classify various CoT techniques, analyzing their principles, trade-offs, and potential for wireless communications. This fundamentals offer a thorough review of how CoT techniques achieve high-quality and reliable reasoning.
    \item We survey the applications of CoT-enhanced LLMs in wireless communications from different key perspectives, including path planning and semantic communication, which highlight the context and advantages of CoT-enabled reasoning across diverse wireless tasks.
    \item We propose a CoT-enhanced LLM framework for intent-driven wireless networks that transforms high-level user intents into control policies via modular Auto-CoT reasoning and DRL-based module activation. A case study of low-altitude UAV networks demonstrate that our framework significantly improves coverage, sum rate, and reasoning quality compared to non-CoT baselines.
\end{itemize}

\section{Overview of Chain-of-Thought}

\subsection{Conventional LLM-empowered Wireless Communications}

LLMs have shown potentials in various wireless communication functions. 

\subsubsection{Intent-driven Networking} LLMs can efficiently interpret high-level user requests and translate them into actionable network configurations by dynamically mapping user intents to specific network parameters such as bandwidth allocation, latency control, and quality of service (QoS) policies.
\subsubsection{Dynamic Route Planning}LLMs can predict optimal data paths by analyzing dynamic traffic patterns, considering parameters such as link quality, congestion levels, and energy efficiency to improve latency and throughput.
\subsubsection{Resource Allocation} LLMs can effectively manage spectrum usage, power control, and user scheduling by continuously analyzing network conditions and adjusting resource distribution to optimize performance under varying loads.

However, despite these capabilities, traditional LLMs often struggle with complex decision-making processes and lack sufficient interpretability and robustness, limiting their effectiveness in dynamic wireless environments. Several reasoning techniques have been developed to improve LLMs' decision-making in dynamic environments. Retrieval-Augmented Generation (RAG) enhances factual grounding by incorporating external documents but is constrained by retrieval quality and domain coverage. Knowledge Graph-based reasoning enables logical inference and interpretability via structured semantics, though graph construction is costly, labor-intensive, and time-consuming. Neuro-symbolic methods integrate neural networks with symbolic rules for robustness and consistency but face challenges in seamless end-to-end integration. In contrast, CoT reasoning encourages step-by-step problem solving within the model, improving reasoning depth with minimal external input. Thus, CoT is especially promising for dynamic and uncertain wireless communication scenarios~\cite{besta2024graph}.

\subsection{Basics of Chain-of-Thought}

CoT has been designed to enhance the reasoning capabilities of LLMs by encouraging step-by-step logical thinking during complex problem-solving tasks~\cite{wei2022chain}. 
Unlike traditional prompting methods, CoT introduces intermediate reasoning steps that explicitly guide the model through the thought process. The key advantages of CoT are summarized as follows:
\begin{itemize}
\item \textbf{Stepwise Reasoning}: CoT can break down complex tasks into structured steps, enhancing interpretability. In intent-driven networks (IDNs), for instance, a high-level intent like ensuring low latency for video streaming can be decomposed into subtasks such as traffic prioritization, routing adjustments, and resource optimization.
\item \textbf{Explainability}: 
CoT improves transparency by explicitly presenting intermediate reasoning. In IDNs, rather than simply tuning parameters, a CoT-based system can explain bandwidth decisions, for example, analyzing traffic types and latency needs, assessing congestion, and determining suitable settings.
\item \textbf{Generalization}: 
By following a structured reasoning path, CoT enhances model adaptability across tasks. An IDN trained for energy efficiency, for example, can generalize to tasks like load balancing or fault tolerance using similar step-by-step logic.

\item \textbf{Mitigating Hallucination}: LLMs may produce outputs that appear plausible but are factually incorrect or physically infeasible~\cite{wang2025wireless}. In intent-driven scenarios like ``optimize uplink power for wearables," CoT helps ensure factual alignment by guiding decisions through grounded, intermediate reasoning that respects constraints such as hardware limits and spectrum regulations.
\end{itemize}

In practice, CoT enhances LLM reasoning by generating intermediate latent representations that model logical dependencies between sub-tasks. Rather than producing an answer in a single pass, the model generates a structured reasoning trace, where each step builds on the previous, narrowing the solution space and reducing reliance on spurious patterns~\cite{zhang2024chain}. This is typically achieved through fine-tuning or prompting with annotated multi-step reasoning paths, reinforcing causal and semantic coherence across tokens. This stepwise structure not only constrains generation to more plausible trajectories but also facilitates intermediate error detection and correction, thereby mitigating hallucinations and improving factual alignment in complex wireless decision-making tasks.

\begin{figure*}
    \centering
    \includegraphics[width=0.99\linewidth]{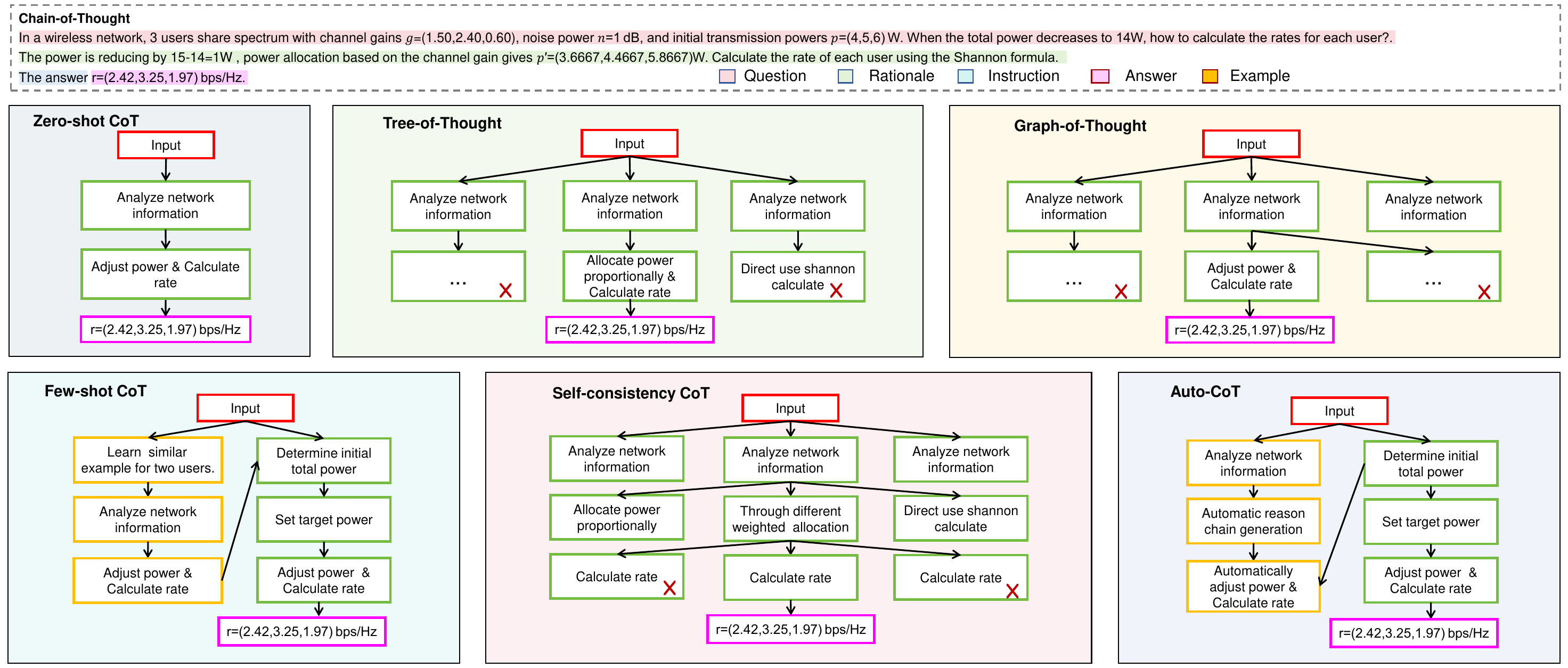}
    \captionsetup{font=scriptsize}
    \caption{An example to illustrate the different chain-of-thought techniques. Zero-shot CoT generates reasoning chains without examples. Few-shot CoT provides exemplars to guide intermediate steps. Self-consistency CoT samples multiple reasoning paths and selects the most consistent answer. Auto-CoT automatically generates exemplars for prompting. Tree-of-Thought explores multiple reasoning branches for better decisions. Graph-of-Thought models reasoning as a revisitable graph structure.}
    \label{Figure1_CoT_model}
\end{figure*}

\subsection{Types of Chain-of-Thought}

As shown in Fig.~\ref{Figure1_CoT_model} we provide an example about IDNs to illustrate the different chain-of-thought techniques.

\subsubsection{Zero-shot CoT} 
Zero-shot CoT activates the step-by-step reasoning capability of LLMs using a simple prompt, typically ``Let's think step by step," without providing examples. This enables logical reasoning in unfamiliar tasks or open-ended questions, including fact verification and commonsense reasoning, such as inferring user intent or estimating bandwidth in dynamic networks~\cite{kojima2022large}. However, its performance degrades on smaller LLMs with fewer than 7 billion parameters.

\subsubsection{Few-shot CoT} 
Few-shot CoT incorporates several exemplars in the prompt, each comprising a problem, reasoning steps, and a correct answer~\cite{kojima2022large}. These examples guide the reasoning process, reducing errors and making it effective for complex tasks like mathematical reasoning and medical diagnosis. However, poor-quality examples can mislead the model, producing incorrect or hallucinated outputs.

\subsubsection{Auto-CoT} 

Auto-CoT is an automated version of Few-shot CoT that leverages clustering techniques or other algorithms to automatically select appropriate examples from a dataset, constructing effective few-shot prompts. Auto-CoT combines the stability of few-shot CoT with the flexibility of zero-shot CoT by automating the sample selection process, minimizing manual intervention. Auto-CoT enhances response accuracy by dynamically retrieving relevant cases, thus is ideal for tasks with recurring user queries such as network troubleshooting or customer support~\cite{zhang2022automatic}. However, suboptimal clustering may yield ineffective prompts.

\subsubsection{Self-consistency CoT}Self-consistency CoT is an enhanced CoT method that combines CoT with a majority voting strategy, generating multiple reasoning paths and selecting the most consistent answer as the final output~\cite{wangself}. This reduces the impact of individual errors and suits tasks with diverse reasoning routes, such as mathematical proofs or network optimization. Since multiple sampling iterations are required, self-consistency CoT incurs a high computational cost.

\subsubsection{Tree-of-Thought} 
Tree-of-Thought (ToT) structures reasoning as a tree, where multiple branches represent different reasoning paths, ultimately converging to the optimal solution~\cite{yao2023tree}. It can dynamically adjust search depth and strategy, excelling in combinatorial optimization and strategic tasks like chess AI, route planning, and congestion control. However, tree-structured reasoning is more complex than linear reasoning, resulting in higher computational overhead. Additionally, without an effective search strategy, the model may encounter computational bottlenecks.

\subsubsection{Graph-of-Thought} 
Graph-of-Thought (GoT) is an advanced CoT-inspired method that leverages graph structures to enhance the reasoning process. It represents reasoning steps as nodes in a directed graph, allowing the model to jump between paths and combine insights from multiple reasoning branches~\cite{besta2024graph}. Therefore, GoT excels in complex scenarios such as pathfinding, social network analysis, and problems with multiple dependencies. However, GoT requires managing a significantly larger number of connections and necessitates additional mechanisms to control the reasoning process to prevent excessive computation.

\begin{figure*}
    \centering
    \includegraphics[width=0.99\linewidth]{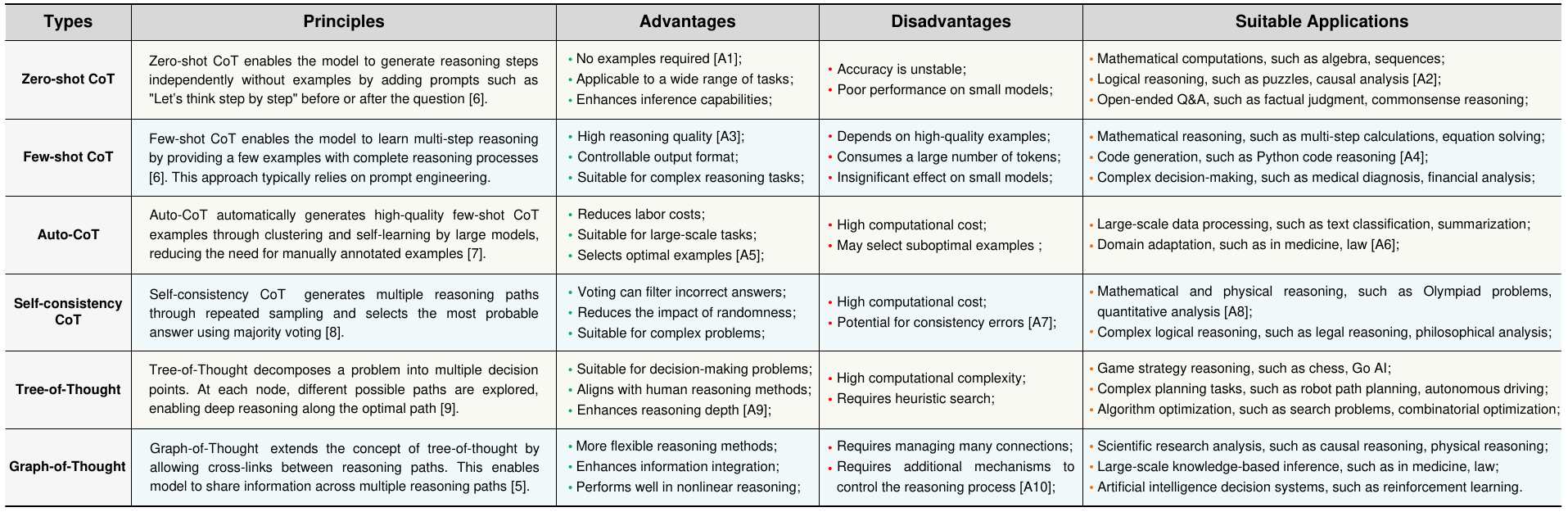}
    \captionsetup{font=scriptsize}
    \caption{The principles, advantages, disadvantages, and suitable applications of different chain-of-thought strategies, including Zero-shot CoT, Few-shot CoT, Auto-CoT, Self-consistency CoT, Tree-of-Thought, and Graph-of-Thought. Related references [A1-A11] and [B1-B4] can be found in \href{https://github.com/wang104225/CoT_Wireless}{{https://github.com/wang104225/CoT\_Wireless}}.}
    \label{Table1_different_CoT}
\end{figure*}

We summarize the different types of chain-of-thought strategies in Fig.~\ref{Table1_different_CoT}, including their principles, advantages, disadvantages, and suitable applications.

\renewcommand{\arraystretch}{1.3}
\setlength{\arrayrulewidth}{0.8pt}

\section{Overview of CoT Applications in Wireless Communications}

In this section, we survey the related work on the applications of CoT-enhanced LLMs in wireless communications, which is summarized in Table.~\ref{tab_summary_of_CoT_in_Wireless}.

\subsection{Wireless-Aware Path Planning}

LLMs, with their natural language and multimodal processing capabilities, offer intuitive, interpretable path planning by enabling dynamic constraint adjustments via user instructions. However, they may generate unrealistic paths due to communication constraint violations or reliance on hallucinated knowledge. The CoT mechanism enhances path reliability and accelerates computation through a phased structured prompting strategy that enforces stepwise reasoning in LLMs. The authors in~\cite{djuhera2024scott} propose a novel Strategic CoT tasking (SCoT) framework, and the path planning process is decomposed into three stages: coarse path generation, region-of-interest refinement, and fine-grained optimization. By using SCoT to reduce the search space of dynamic programming, the proposed path planning system achieves up to $62\%$ reduction in computation time while maintaining near-optimal path gain.

\vspace{-0.3cm}
\subsection{Resource Allocation}

LLMs are being adopted in wireless communications for resource allocation tasks such as scheduling and network slicing, yet they struggle with multimodal inputs, task decomposition, and adaptability in dynamic settings. Their tendency toward hallucination can cause critical orchestration errors, such as misallocating frequency bands based on fabricated interference patterns, resulting in inefficient spectrum use. CoT helps LLMs decompose complex resource allocation problems into sub-tasks, quantify intermediate results, support dynamic optimization, reduce resource waste, and improve performance. In~\cite{10930397}, CoT is applied in the planning module of the proposed WirelessAgent framework by prompting the LLM to iteratively reason through network slicing decisions, such as evaluating traffic load, matching user QoS intents with slice templates, and selecting optimal configurations. Simulation results show that incorporating CoT reasoning in planning enables interpretable, fine-grained decisions and achieves a $12.4\%$ reduction in power consumption and an $18.7\%$ improvement in early-stage service quality.

\vspace{-0.3cm}
\subsection{Semantic Communications}

In semantic communication, LLMs enhance efficiency by acting as shared knowledge bases (SKBs) for semantic compression and generation. However, LLMs struggle with task decomposition, semantic ambiguity, and adapting to dynamic channels, particularly when prioritizing text semantics over image generation. These limitations hinder multi-user collaboration, decision efficiency, and robustness in real-time wireless semantic scenarios. CoT addresses this by structuring outputs for hierarchical semantic parsing and adaptive sequential extraction, improving channel adaptability. In~\cite{yang2024rethinking}, a multi-user generative semantic communication framework (M-GSC) is proposed, where an LLM-based SKB performs Zero-shot CoT-driven task decomposition, semantic specification, and mapping. By leveraging Zero-shot CoT for complex task decomposition and representation selection, the framework enables efficient semantic decoding offloading. The case study shows that M-GSC outperforms SwinJSCC benchmark by up to $38.7\%$ in semantic accuracy at 0 dB SNR.

\vspace{-0.3cm}
\subsection{Integrated Sensing and Communication}

\begin{table*}[htp] \scriptsize
  \centering
  \captionsetup{font=scriptsize}
  \caption{Summary of Chain-of-Thought for LLM-empowered Wireless Communications and Networking.}
  \label{tab_summary_of_CoT_in_Wireless}
    \begin{tabular}{m{0.11\textwidth}<{\centering}||m{0.06\textwidth}<{\centering}|m{0.3\textwidth}<{\centering}|m{0.43\textwidth}<{\centering}}
      \hline
      \textbf{Types}  &  \textbf{Reference} & \textbf{Features} & \textbf{ \textcolor{green}{\ding{51}} Advantages \& \textcolor{red}
          {\ding{115}} Future works} \\
       \hline
       
      Wireless-Aware Path Planning &\cite{djuhera2024scott} 
      & A novel path planning strategy integrating AI-driven optimization with real-time environmental adaptation 
      & \begin{itemize}[leftmargin=*] 
        
          \item[\textcolor{green}{\ding{51}}] Higher computational efficiency 
          \hfill \textcolor{green}{\ding{51}} Enhanced path reliability
          \item[\textcolor{green}{\ding{51}}] Improved adaptability to dynamic environments
          \item[\textcolor{red}
          {\ding{115}}] Build CoT verification module
          \hfill \textcolor{red}{\ding{115}} Dynamic CoT online updating
      
      \vspace{-1.0em}
      \end{itemize}\\
      \hline
      
      \multirow{2}{0.11\textwidth}[-25pt]{\centering Resource Allocation} 
      & \cite{10930397} 
      & A prompt engineering framework for optimizing LLM applications in wireless networks  
      & \begin{itemize}[leftmargin=*]

          \item[\textcolor{green}{\ding{51}}] Efficient inference with low computational overhead
          \item[\textcolor{green}{\ding{51}}] Rapid deployment to dynamic network environments
          \item[\textcolor{green}{\ding{51}}] Avoidance of complex model training and fine-tuning.
          \item[\textcolor{red}{\ding{115}}] Design dynamic adjustment mechanism for CoT
      \vspace{-1.0em}
      \end{itemize}\\
      \cline{2-4}
      & B-1 & Autonomous resource orchestration using LLMs and RL for dynamic network environments 
      & \begin{itemize}[leftmargin=*]
      
          \item[\textcolor{green}{\ding{51}}] Enhanced reasoning and planning capabilities
          \item[\textcolor{green}{\ding{51}}] Continuous adaptation to network dynamics
          \item[\textcolor{green}{\ding{51}}] Reduced computational overhead through efficient resource management
          \item[\textcolor{red}{\ding{115}}] Optimize CoT-RL collaboration mechanism
      \vspace{-1.0em}
      \end{itemize}\\
      \cline{2-4}

      & B-2 
      & A joint caching and inference framework for sustainable LLM agent services in SAGINs 
      & \begin{itemize}[leftmargin=*]
      
          \item[\textcolor{green}{\ding{51}}] Efficient resource utilization through model caching
          \item[\textcolor{green}{\ding{51}}] Enhanced service quality with reduced latency and improved accuracy
          \item[\textcolor{green}{\ding{51}}] Scalability and adaptability to complex network environments
          \item[\textcolor{red}{\ding{115}}] Design CoT step suppression mechanism
      \vspace{-1.0em}
      \end{itemize}\\

      \hline
      
      \multirow{4}{0.11\textwidth}[+5pt]{\centering Semantic Communications} 
      &  \cite{yang2024rethinking}  
      & Distributed Semantic Communication Framework with SKB  
      & \begin{itemize}[leftmargin=*]

          \item[\textcolor{green}{\ding{51}}] Semantic encoding standardization
          \hfill \textcolor{green}{\ding{51}} Efficient resource management
          \item[\textcolor{green}{\ding{51}}] Personalized semantic decoding
          \hfill \textcolor{red}{\ding{115}} Multimodal CoT deep integration
    
      \vspace{-1.0em}
      \end{itemize}\\
      \cline{2-4}
      
      & B-3 
      & Semantic-Driven Multi-Agent Communication with Reduced Signaling Expenditure 
      & \begin{itemize}[leftmargin=*]
      
          \item[\textcolor{green}{\ding{51}}] Lower communication overhead
          \hfill \textcolor{green}{\ding{51}} Faster convergence speed
          \item[\textcolor{green}{\ding{51}}] Improved decision-making accuracy
          \hfill \textcolor{red}{\ding{115}} Adaptive CoT length control
      \vspace{-1.0em}
      \end{itemize}\\
      
      
      
      
      \hline
      \multirow{4}{0.11\textwidth}[-1pt]
      {\centering Integrated Sensing and Communication} 
      &\cite{10829757} 
      & A novel path planning strategy integrating AI-driven optimization with real-time environmental adaptation 
      & \begin{itemize}[leftmargin=*] 
           
          \item[\textcolor{green}{\ding{51}}] Higher computational efficiency 
          \hfill \textcolor{green}{\ding{51}} Enhanced path reliability
          \item[\textcolor{green}{\ding{51}}] Improved adaptability to dynamic environments
          \item[\textcolor{red}{\ding{115}}] Balance semantic conflicts across modalities
      
      \vspace{-1.0em}
      \end{itemize}\\
      \cline{2-4}
      
      &  B-4 
      & A split learning system integrating ISAC and CoT for efficient LLM agent deployment in 6G networks.  
      & \begin{itemize}[leftmargin=*]
      
          \item[\textcolor{green}{\ding{51}}] Efficient resource utilization through end-edge-cloud computing
          \item[\textcolor{green}{\ding{51}}] Enhanced adaptability to dynamic environments
          \item[\textcolor{green}{\ding{51}}] Improved decision-making with reliable reasoning and planning
          \item[\textcolor{red}{\ding{115}}] Optimize hierarchical system CoT allocation
      
      
            
      

            
      
      \vspace{-1.0em}
      \end{itemize}\\
      \hline
    \end{tabular}
\end{table*}

LLMs, through multimodal instruction tuning and cross-modal reasoning, have been successfully applied to ISAC tasks such as beam prediction and obstacle detection, enhancing sensing-communication coordination.
However, LLM-assisted ISAC faces goal misalignment and substantial spatiotemporal differences across modalities.
CoT can address these by structuring instructions to filter irrelevant semantics, model spatiotemporal continuity, and enable dynamic cross-modal feature extraction. 
In~\cite{10829757}, an MLLM-based end-to-end ISAC architecture employs the CoT mechanism with structured instructions and distributed inference to decompose complex tasks like beam prediction, progressively associating the semantics of multimodal data, e.g., identifying obstacles in images and analyzing target interference in radar data. This approach achieves Top-3 beam prediction accuracies of $70\%$ and $95\%$ on two ISAC tasks, surpassing baseline methods.

\vspace{-0.3cm}
\subsection{Lessons Learned}

The above cases highlight that the CoT reasoning mechanism can adapt to dynamic environmental changes and enhance decision-making accuracy, while wireless communication networks are inherently subject to complex and constantly evolving conditions. Therefore, we believe that integrating CoT into wireless network management can substantially improve adaptability and reliability.

\section{Proposed CoT Framework for Wireless Communications}

\subsection{The Proposed Framework}

\begin{figure*}
    \centering
    \includegraphics[width=0.99\linewidth]{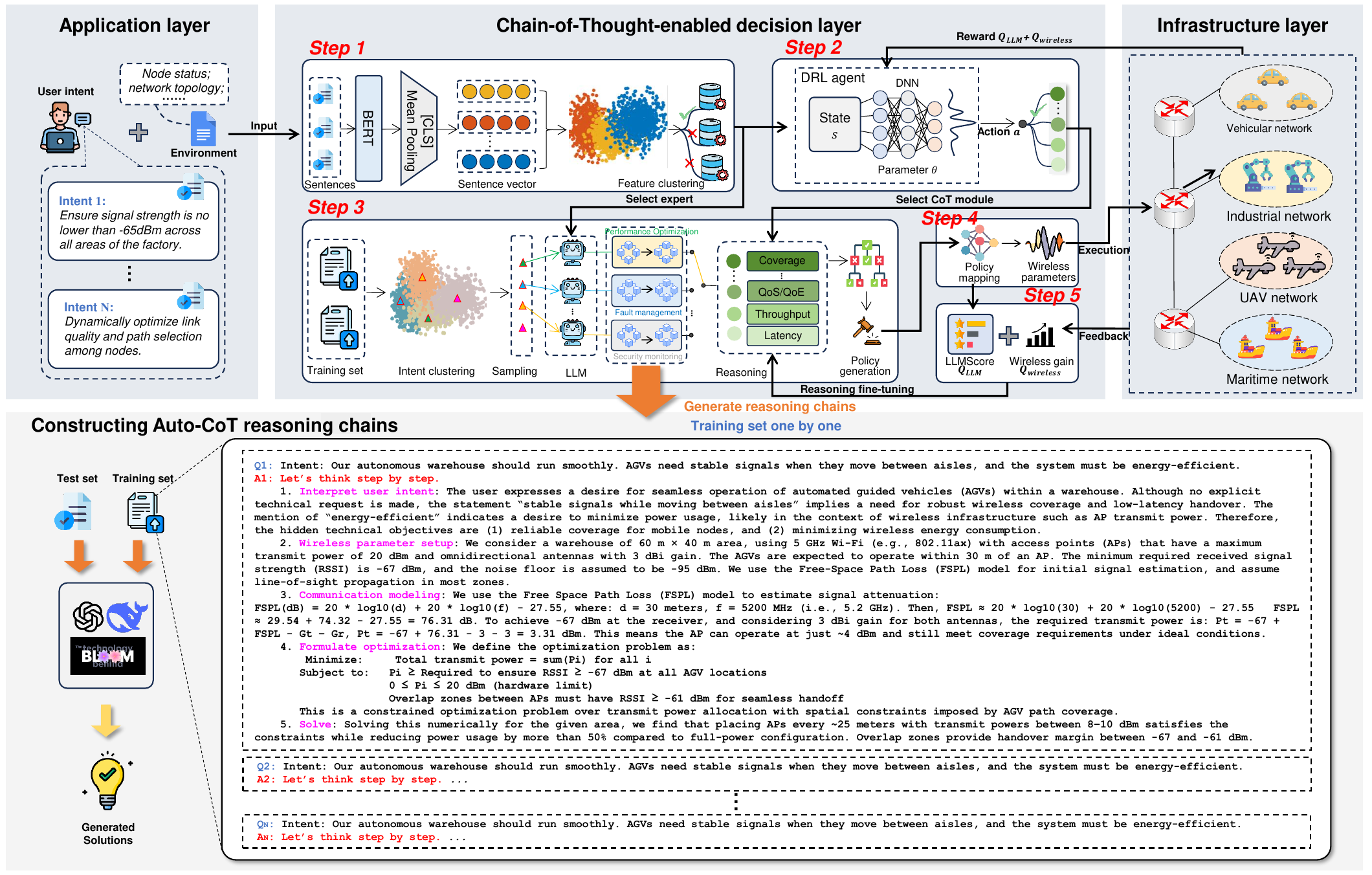}
    \captionsetup{font=scriptsize}
    \caption{The proposed CoT-enabled IDNs framework operates across three layers: the application layer, the Chain-of-Thought-enabled decision layer, and the infrastructure layer. In the application layer, diverse user intents and environmental information are collected and encoded. In the Chain-of-Thought-enabled decision layer, intent understanding and decision generation are achieved through \textbf{Steps 1–5}. In addition, a case of Auto-CoT reasoning chain construction is presented. In the infrastructure layer, the generated policies are then executed in real-world network environments.}
    \label{framework}
\end{figure*}

To enable efficient, interpretable, and adaptive wireless optimization, we propose a CoT-enabled multi-layer framework for intent-driven networks that translates natural language intents into executable control strategies. As shown in Fig.~\ref{framework}, the framework comprises three layers: the application layer captures user intents and environmental data; the CoT-enabled decision layer performs intent understanding and strategy generation through Steps 1–5; and the infrastructure layer enacts these strategies in real-world wireless environments. Detailed descriptions of the steps follow.

\subsubsection{\textbf{Step 1. Intent Parsing and Clustering}} 
First, high-level intents expressed in natural language are sent to the server and are then encoded into dense vector representations using the Sentence-BERT language embedding technique, capturing contextual semantics of user intents, such as performance optimization, fault management, or security monitoring. Next, we apply K-Means clustering to group similar intents, thereby partitioning the global task space into tractable sub-domains.

\subsubsection{\textbf{Step 2. Intent-Aware Module Activation}} 
After categorizing the user request, a deep reinforcement learning (DRL) agent dynamically selects the most suitable CoT reasoning module based on the request type and real-time system states, including target of optimization, user-specific QoS requirements and resource constraints. 
Trained via a deep Q-network (DQN), the DRL agent optimizes a reward function that balances performance gains against inference quality, and the definition of this reward function can be found in Step 5.

\subsubsection{\textbf{Step 3. CoT Reasoning}} 
Each selected module encapsulates task-specific CoT examples aligned with the classified intent. As shown in Fig. 3, a module for power allocation guides the LLM to reason step-by-step through communication modeling, regulatory constraints, and optimization objectives before deriving a power control policy. These modules incorporate few-shot exemplars to steer the LLM toward structured, logical inferences. By decomposing complex objectives into intermediate reasoning steps, the modular CoT design enhances both interpretability and robustness.

\subsubsection{\textbf{Step 4. Intent Realization}} 
After CoT reasoning, the generated natural language strategies are parsed into executable network control commands using a strategy extraction engine based on neural semantic parsing. These commands map abstract reasoning to concrete parameters such as transmit power, resource blocks, or scheduling actions. The derived strategies are then delivered to edge nodes or user terminals for real-time execution. Execution feedback, such as interference quality and actual throughput, can be relayed back to the central server, forming a closed-loop interaction that supports continual fine-tuning of the LLM reasoning modules to improve the reasoning process.

\subsubsection{\textbf{Step 5. Joint Metric Evaluation}} 
To evaluate how well user intents are fulfilled, a multi-dimensional evaluation module is introduced, combining both communication-layer and reasoning-layer metrics. Traditional physical-layer metrics such as sum rate and energy consumption are used to quantify the real-world performance of the deployed strategies. In parallel, the fitness score proposed in~\cite{li2025evolving} is used to evaluate the quality of LLM-generated outputs, which includes consistency score, informativeness, and misleadingness score. These aspects are unified into a composite utility function, capturing both system-level impact and reasoning fidelity. The composite utility function is finally formulated as $Q_{total}=\alpha Q_{LLM}+\beta Q_{wireless}$, where $Q_{LLM}$ represent inference quality, $Q_{wireless}$ is wireless performance gain, $\alpha$ and $\beta$ are weighting parameters.

\subsection{Case Study: Intent-driven Deployment Optimization and Power Control for UAV networks}


To evaluate the proposed CoT-based framework, we conduct a case study involving UAV deployment and transmit power control in an intent-driven network. The network administrator expresses a high-level intent: ``Deploy a UAV base station that maximizes both coverage and user data rates.'' The system translates this abstract intent into concrete wireless decisions involving UAV deployment and transmit power allocation. We adopt the Auto-CoT strategy to construct multi-step reasoning chains based on this intent. An LLM is prompted with domain-specific exemplars and wireless environment configurations, including user locations and channel models.

\textbf{Experimental Setup:} The environment is modeled as a $1,000~\mathrm{m} \times 1,000~\mathrm{m}$ area with users randomly distributed according to a uniform spatial process. 
We set the carrier frequency to $2.4~\mathrm{GHz}$, and the system bandwidth to $20~\mathrm{MHz}$. We consider the free space path loss (FSPL) as the path loss model. Besides, the UAV is fixed at an altitude of $100~\mathrm{m}$, with a maximum transmit power of $20~\mathrm{dBm}$. Noise power is computed based on the Boltzmann constant and ambient temperature of $290~\mathrm{K}$. The communication range of the UAV is set to $200\mathrm{-}600~\mathrm{m}$. As for LLMs, GPT-3.5 and GPT-4o are adopted from OpenAI due to their fast inference speed and improved stability in generating consistent CoT reasoning chains, where these models are not fine-tuned but are instead guided by carefully designed CoT prompts and task-specific exemplars. All experiments are conducted on a Ubuntu desktop equipped with 16 GB RAM and an AMD Ryzen 7 6800H CPU operating at 3.2 GHz, using Python 3.12. Note that our experimental results are based on the average of $10$ random samples of user locations.

\textbf{Problem-Solving Process:} The CoT-based reasoning proceeds in structured steps: 1) parse and translate the user's high-level intent into quantifiable goals, i.e., coverage maximization and sum-rate improvement; 2) extract user coordinates and wireless parameters from inputs; 3) formulate a multi-objective optimization using FSPL-based SINR and Shannon theory; 4) solve for UAV positions and transmit powers via heuristic algorithms under coverage and interference constraints; and 5) compute a composite utility combining coverage ratio, sum rate, and CoT response quality. 
A composite utility function is used as the optimization criterion, jointly accounting for the coverage ratio, aggregate sum rate, and LLM-generated reasoning quality. 
The final utility function is set to
$Q_{\mathrm{total}} = \alpha Q_{\mathrm{LLM}} + \beta \left(Q_{c}+Q_{R}\right)$, where $\alpha=0.1$, $\beta=0.45$, $Q_c$ denotes the coverage rate, and $Q_R$ represents the normalized sum rate. Through CoT-enabled reasoning, LLMs autonomously explore trade-offs between UAV location, and SINR-based sum rate to derive an effective strategy. 

\vspace{-0.3cm}
\subsection{Experimental Results}

\begin{figure}[!t]
    \centering
    \includegraphics[width=0.7\linewidth]{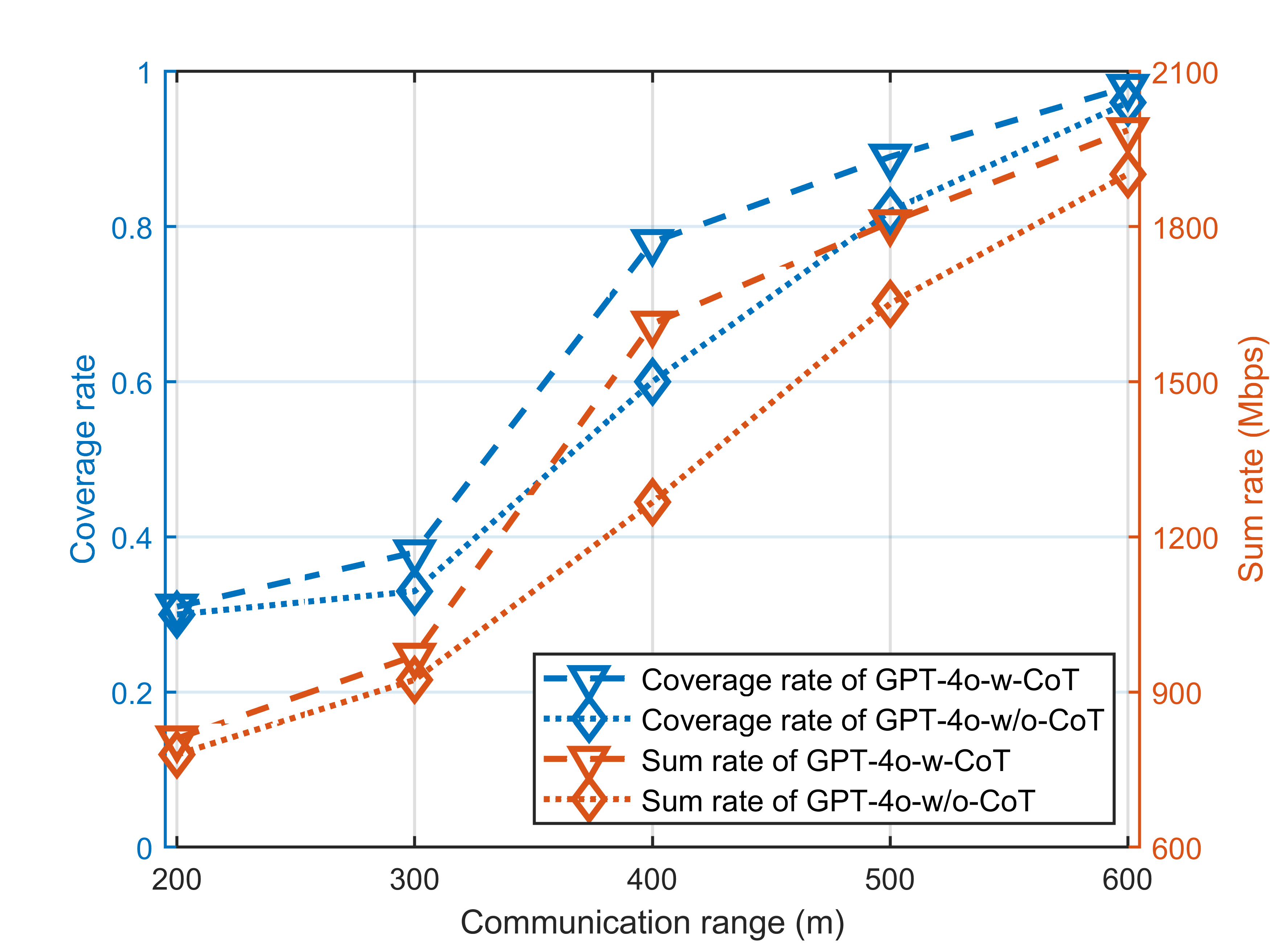}
    \captionsetup{font=scriptsize}
    \caption{Coverage rate and sum rate performance of the proposed framework versus the communication range of UAV under two configurations: GPT-4o with CoT (\textbf{GPT-4o-w-CoT}), and GPT-4o without CoT (\textbf{GPT-4o-w/o-CoT}), and all employ DRL-enhanced module activation.}
    \label{sumrate_coverage_vs_communication_range}
\end{figure}

Fig.~\ref{sumrate_coverage_vs_communication_range} compares the coverage and sum rate performance of intent-driven framework with and without CoT reasoning across different UAV communication ranges. As the communication range increases from $200~\mathrm{m}$ to $550~\mathrm{m}$, both metrics exhibit significant improvement due to enhanced signal reach and user association flexibility. Notably, the CoT-enabled IDN consistently outperforms the non-CoT baseline across all ranges. In particular, the CoT-based approach achieves approximately a $27.2\%$ increase in sum rate compared to the non-CoT approach at $400~\mathrm{m}$. These gains stem from the structured reasoning process enabled by Auto-CoT, which helps the LLM accurately decompose user intent, optimize UAV positioning, and assign transmit power while accounting for SINR degradation and interference. In contrast, the baseline model lacks explicit intermediate reasoning steps, leading to suboptimal placement and inefficient resource utilization. This is because CoT enables step-by-step intent interpretation and decision refinement, allowing the model to better handle complex constraints and coordination challenges in UAV networks.

\begin{figure}[!t]
    \centering
    \includegraphics[width=0.97\linewidth]{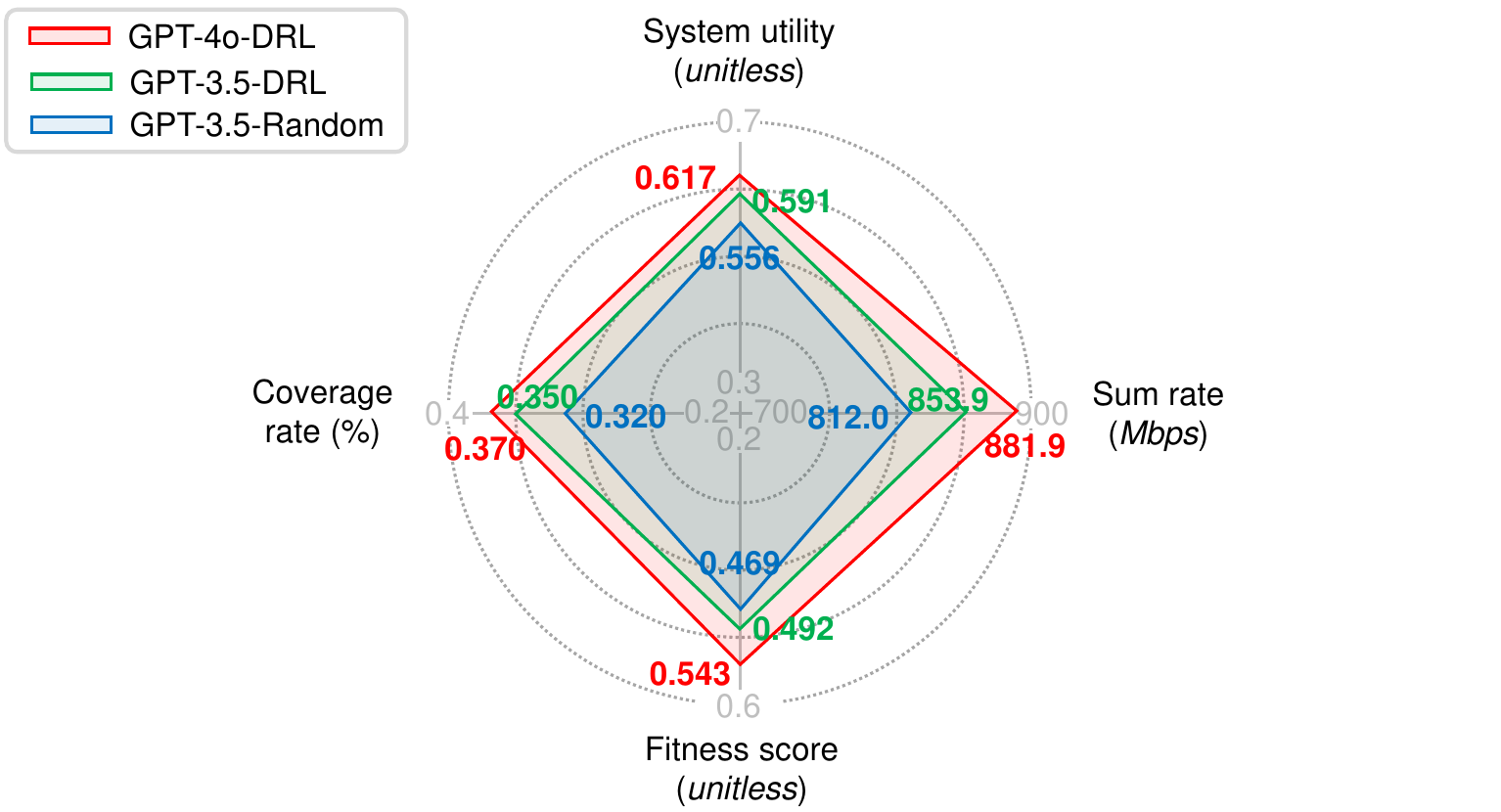}
    \captionsetup{font=scriptsize}
    \caption{Comprehensive performance comparison of the proposed CoT-driven framework under three configurations: CoT-enhanced GPT-4o with DRL activation (\textbf{GPT-4o-DRL}), CoT-enhanced GPT-3.5 with DRL activation (\textbf{GPT-3.5-DRL}), and CoT-enhanced GPT-3.5 with random activation (\textbf{GPT-3.5-Random}).}
    \label{performance_of_different_LLM_final}
\end{figure}

Fig.~\ref{performance_of_different_LLM_final} presents a comprehensive performance comparison of the proposed CoT-driven framework under three configurations: GPT-4o-DRL, GPT-3.5-DRL, and GPT-3.5-Random. Across all metrics, GPT-4o-DRL outperforms GPT-3.5-DRL, demonstrating that more advanced LLMs enhance reasoning capabilities, resulting in more accurate and coherent multi-step CoT chains. Additionally, GPT-3.5-DRL consistently outperforms GPT-3.5-Random, indicating that DRL-based module activation can intelligently select the most suitable CoT construction. In contrast, random activation fails to guide the CoT module toward reasoning paths aligned with task-specific requirements, often leading to suboptimal and unstructured decision outputs.

\section{CONCLUSION}

This paper has explored the integration of CoT into LLM-driven wireless communications. We have first reviewed the foundational CoT techniques and their advantages. Practical applications have been analyzed to highlight the interpretability and effectiveness of CoT-enabled wireless communications, such as semantic communications and ISAC. Furthermore, we have proposed the CoT-enhanced LLMs framework for intent-driven networks that translates user intents into control policies. A UAV deployment case study has demonstrated notable gains in communication performance and reasoning quality over non-CoT baselines. Future work can focus on enhancing CoT interpretability and enabling federated reasoning at the edge for greater scalability and intelligence.

\bibliographystyle{IEEEtran}
\bibliography{references}



\end{document}